\documentstyle{article}

\begin{document}

\begin{center}
{\large {\bf Angular momentum and energy structure of the coherent state of
a 2D isotropic harmonic oscillator}}

 HUO Wujun, LIU Yufeng and ZENG Jinyan

Department of Physics, Peking University, Beijing, 100871, P. R. China
\end{center}

\noindent{\large {\bf Abstract}}

The angular momentum structure and energy structure of the coherent state of
a 2D isotropic harmonic oscillator were investigated. Calculations showed
that the average values of angular momentum and energy (except the zero
point energy) of this nonspreading 2D wave packet are identical to those of
the corresponding classical oscillator moving along a circular or an
elliptic orbit.
\begin{center}
{\large {\bf I. Introduction}}
\end{center}

The coherent state of a harmonic oscillator was first constructed by
Schr\"odinger$^{[1,2]}$ and since the sixties was widely used in the
description of coherent light sources and in communication theory at optical
frequency$^{[3,4]}$. The main motivation of Schr\"odinger was to investigate
the relation between quantum mechanics and classical mechanics$^{[5]}$. His
aim was to find a special kind of quantum state --- a nonspreading wave
packet whose center follows the corresponding classical motion. He believed
that it is only a question of computational skill to accomplish the same
thing for electron in the hydrogen atom. However, wave packets describing
the Kepler orbits in a hydrogen atom are yet to be discovered, which is
usually considered to be connected with the nonuniformity of the hydrogen
spectrum$^{[6]}$. Therefore, someone tried to find the wave packets 
constructed by
the superposition of Rydberg's states$^{[7]}$. Nieto and Simmons have
constructed approximate (not exact) coherent states for particle in general
one-dimensional (1D) potentials$^{[8,9]}$.

In classical mechanics a 2D isotropic harmonic oscillator follows, in
general, an elliptic orbit, which is reduced to a circular orbit or a
straight line in special cases. It is expected that the coherent states of a
2D isotropic harmonic oscillator are nondispersing wave packets with centers
moving along elliptic orbits. However, as we know, maybe due to
computational difficulties, the classical correspondence (angular momentum
structure, energy structure) of such coherent states have not been
investigated in detail . In this letter the angular momentum and energy
constituents of such 2D nonspreading wave packets were calculated and it was
shown that the centers of the wave-packets follow the identical elliptic
orbits as the corresponding 2D classical oscillator.

\begin{center}
{\large {\bf II. Angular momentum and energy constituents of the coherent
state of a 2D isotropic harmonic oscillator}}
\end{center}

The Schr\"odinger's coherent state of a 1D harmonic oscillator is well-known$%
^{[1,2]}$, 
\begin{equation}
\label{1} 
\begin{array}{c}
\psi _{\xi _0}(\xi ,t)= 
\frac{\alpha ^{1/2}}{\pi ^{1/4}}\exp [-\frac{i\omega t}2-\frac 12\xi
^2-\frac 14\xi _0^2(1+e^{2i\omega t})+\xi _0\xi e^{-i\omega t}], 
\end{array}
\end{equation}
where $\xi
=\alpha x,\xi _0=\alpha x_0,\alpha = 
\sqrt{M\omega /\hbar }$, and $\left| \psi \right| ^2=\frac \alpha {\sqrt{\pi }%
}\exp [-(\xi -\xi _0\cos \omega t)^2]$. 

The shape of this wave packet remains unchanged as time progresses and the
position of its center is located at $\xi =\xi _0\cos \omega t$, which is
the same as the motion of a classical oscillator with amplitude $x_0=\xi
_0/\alpha $ and natural angular frequency $\omega $.

Assume the phase of the coherent state along the $y$ direction be $\pi /2$
retarded with respect to that along the $x$ direction, 
\begin{equation}
\label{2} 
\begin{array}{c}
\psi _{\eta _{_0}}(\eta ,t)= 
\frac{\alpha ^{1/2}}{\pi ^{1/4}}\exp [-\frac{i(\omega t-\pi /2)}2-\frac
12\eta ^2-\frac 14\eta _0^2(1-e^{2i\omega t})+i\eta _0\eta e^{-i\omega t}],
\\ \eta =\alpha y\ ,\ \eta _0=\alpha y_{0,} 
\end{array}
\end{equation}
whose center is located at $\eta =\eta _0\cos (\omega t-\pi /2)$. Thus the
coherent state of a 2D isotropic harmonic oscillator is 
\begin{equation}
\label{3} 
\begin{array}{c}
\psi _{\xi _0\eta _{_0}}(\xi ,\eta ,t)=\frac \alpha {\pi ^{1/2}}\exp
[-i\omega t+i\frac \pi 4-\frac 12(\xi ^2+\eta ^2)-\frac 14\xi
_0^2(1+e^{2i\omega t})-\frac 14\eta _0^2(1-e^{2i\omega t}) \\ 
-(\xi _0\xi +i\eta _0\eta )e^{-i\omega t}], 
\end{array}
\end{equation}
The wave function at initial time ($t=0$) is ( the trivial constant phase
factor $e^{i\pi /4}$ being neglected). 
\begin{equation}
\label{4}\psi _c(\xi ,\eta )=\frac \alpha {\pi ^{1/2}}\exp [-\frac 12(\xi
^2+\eta ^2)-\frac 12\xi _0^2+(\xi _0\xi +i\eta _0\eta )]. 
\end{equation}
This 2D coherent state is a nonstationary state which is a coherent
superposition of infinite of stationary states. To investigate its angular
momentum structure and energy structure, we may expand (4) in terms of the
simultaneous eigenstates of the complete set of conserved quantities ($H,l_z$%
), and the moduli of the expansion coefficients are time-independent. The
normalized simultaneous eigenstates of ($\hat H,\hat l_z$) for a 2D
isotropic oscillator may be expressed as 
\begin{equation}
\label{5} 
\begin{array}{c}
\psi _{mn_r}(\tilde \rho ,\varphi )=\left[ 
\frac{n_r!\alpha ^2}{\pi (\left| m\right| +n_r)!}\right] ^{1/2}e^{im\varphi
}\tilde \rho ^{\left| m\right| }e^{-\tilde \rho ^2/2}L_{n_r}^{\left|
m\right| }(\tilde \rho ^2)\ , \\ n_r\ ,\;\left| m\right| =0,1,2,\cdots
,\quad \tilde \rho =\alpha \rho =\alpha \sqrt{x^2+y^2}=\sqrt{\xi ^2+\eta ^2}%
, 
\end{array}
\end{equation}
where $L$ is the generalized Laguerre polynomial$^{[9]}$, and the
corresponding eigenvalue is 
\begin{equation}
\label{6}E=E_N=(N+1)\hbar \omega \ ,\quad N=2n_r+\left| m\right|
=0,1,2,\cdots 
\end{equation}
The expansion coefficients of $\psi _c$ in terms of $\psi _{mn_r}$are 
\begin{equation}
\label{7}C_{mn_r}=\int_0^{2\pi }d\varphi \int_0^\infty \rho d\rho \psi
_c(\xi ,\eta )\psi _{mn_r}^{*}(\tilde \rho ,\varphi ), 
\end{equation}
which can be calculated in two cases:

I. $\xi _0=\eta _0$ ( circular orbit)

Substituting (4) and (5)\ into (7), careful calculation (Appendix) shows
that 
\begin{equation}
\label{8}C_{mn_r}=\left\{ 
\begin{array}{c}
\xi _0^me^{-\xi _0^2/2}(\frac 1{m!})^{1/2}\delta _{n_r0}\ ,\quad \ (m\geq 0)
\\ 
0\qquad \quad \qquad \qquad \quad \ ,\quad (m<0) 
\end{array}
\right. 
\end{equation}
This is expected because the quantum state corresponding to a classical
circular orbit must have $n_r=0$ ( radial wave function without node). $%
m\geq 0$ in (8) means that the oscillator moves counter-clockwise along a
circular orbit. If the phase of the coherent state along the $y$ direction
is $\pi /2$ advanced with respect to that along the $x$ direction, $C_{mn_r}$
does not vanish only for $m<0$, which means that the circular motion is
clockwise.

Using (8) we may investigate the angular momentum structure and energy
structure of the 2D coherent state (3). First, the average value of $m$ is 
\begin{equation}
\label{9}\overline{m}=\sum_{m=0}^\infty m\xi _0^{2m}e^{-\xi _0^2}\frac
1{m!}=\xi _0^2\ , 
\end{equation}
hence, the average value of angular momentum $l_z$ is 
\begin{equation}
\label{10}\overline{l_z}=\overline{m}\hbar =\xi _0\hbar =M\omega
x_0^2=MR^2\omega \ , 
\end{equation}
where $R=x_0$ is the radius of circular orbit. it is seen that $\overline{l_z%
}$ is the same as the angular momentum of the corresponding classical 2D
oscillator moving along a circular orbit with radius $R$ and angular
frequency $\omega $. Second, we may calculate the average value of energy
using (8) (note: $n_r=0,N=\left| m\right| $) 
\begin{equation}
\label{11}\overline{H}=(\overline{m}+1)\hbar \omega =MR^2\omega +\hbar
\omega \ . 
\end{equation}
It is seen that $\overline{H}$ ( except the zero-point energy $\hbar \omega $%
) is just the energy of the corresponding classical oscillator moving along
a circular orbit with radius $R$ and angular frequency $\omega $.

II. \ $\xi _0\neq \eta _0$ (elliptic orbit)

Let 
\begin{equation}
\label{12}A=(\xi _0-\eta _0)/2\ ,\quad B=(\xi _0+\eta _0)/2\ , 
\end{equation}
calculation (Appendix) shows that 
\begin{equation}
\label{13}C_{mn_r}=\left\{ 
\begin{array}{c}
(-1)^{n_r}\left[ \frac 1{n_r!(m+n_r)!}\right] ^{1/2}e^{-\xi
_0^2/2}e^{AB}A^{n_r}B^{m+n_r}\quad ,(m\geq 0), \\ 
(-1)^{n_r}\left[ \frac 1{n_r!(-m+n_r)!}\right] ^{1/2}e^{-\xi
_0^2/2}e^{AB}B^{n_r}A^{-m+n_r}\ ,(m<0)\,. 
\end{array}
\right. 
\end{equation}
Using (13) we may calculate 
\begin{equation}
\label{14}\overline{m}(m\geq 0)=\sum_{m\geq 0,n_r}\left| C_{mn_r}\right|
^2m\ ,\quad \overline{m}(m<0)=\sum_{m<0,n_r}\left| C_{mn_r}\right| ^2m\ , 
\end{equation}
\begin{equation}
\label{15}\overline{n_r}(m\geq 0)=\sum_{m\geq 0,n_r}\left| C_{mn_r}\right|
^2n_r\ ,\quad \overline{n_r}(m<0)=\sum_{m<0,n_r}\left| C_{mn_r}\right|
^2n_r\ . 
\end{equation}
For example, 
\begin{equation}
\label{16} 
\begin{array}{c}
\overline{n_r}(m\geq 0)=e^{-\xi _0^2}e^{2AB}[\frac{A^2}{1!}(e^{B^2}-1)+2 
\frac{A^4}{2!}(e^{B^2}-1-\frac{B^2}{1!})+3\frac{A^6}{3!}(e^{B^2}-1-\frac{B^2 
}{1!}-\frac{B^4}{2!})+\cdots ] \\ \qquad \ \ \qquad =A^2e^{-\xi
_0^2+A^2+2AB+B^2}-e^{-\xi _0^2+2AB}[\frac{A^2}{1!}+2\frac{A^4}{2!}(1+\frac{%
B^2}{1!})+3\frac{A^6}{3!}(1+\frac{B^2}{1!}+\frac{B^4}{2!})+\cdots ]\ . 
\end{array}
\end{equation}
Similarly, it can be shown that 
\begin{equation}
\label{17} 
\begin{array}{c}
\overline{(-m+n_r)}(m<0)=e^{-\xi _0^2+2AB}[1\cdot (1\frac{A^2}{1!}+2\frac{%
A^4 }{2!}+\cdots ) \\ \qquad \qquad \qquad \qquad \qquad \qquad + 
\frac{B^2}{1!}(2\frac{A^4}{2!}+3\frac{A^6}{3!}+\cdots ) \\ \qquad \qquad
\qquad \qquad \qquad \qquad + 
\frac{B^2}{2!}(3\frac{A^6}{3!}+4\frac{A^8}{4!}+\cdots ) \\ \;\qquad \qquad
\quad +\cdots ]\ . 
\end{array}
\end{equation}
Hence, we get 
\begin{equation}
\label{18}\overline{n_r}(m\geq 0)+\overline{(-m+n_r)}(m<0)=A^2e^{-\xi
_0^2+A^2+2AB+B^2}=A^2\ . 
\end{equation}
Similarly, 
\begin{equation}
\label{19}\overline{n_r}(m<0)+\overline{(m+n_r)}(m\geq 0)=B^{2.} 
\end{equation}
(19)$\pm $(18) result in, respectively, 
\begin{equation}
\label{20}2\overline{n_r}+\left| \overline{m}\right| =A^2+B^2, 
\end{equation}
\begin{equation}
\label{21}\overline{m}=B^2-A^2. 
\end{equation}
Therefore, we get 
\begin{equation}
\label{22}\overline{l_z}=\overline{m}\hbar =(B^2-A^2)\hbar =\xi _0\eta
_0\hbar =x_0y_0M\omega 
\end{equation}
which is just the angular momentum of a classical oscillator moving along an
elliptic orbit with semi-major and semi-minor axes of $x_0$ and $y_0$. The
average value of energy is 
\begin{equation}
\label{23} 
\begin{array}{c}
\overline{H}=\overline{(2n_r+\left| m\right| +1)}\hbar \omega \\ \;\;\
=(A^2+B^2)\hbar \omega +\hbar \omega \\ 
\quad =\frac 12(\xi _0^2+\eta _0^2)\hbar \omega +\hbar \omega \\ 
\quad \quad =\frac 12(x_0^2+y_0^2)M\omega ^2+\hbar \omega 
\end{array}
\end{equation}
which is also the same as that of a classical oscillator moving along an
elliptic orbit (except the zero-point energy $\hbar \omega $).

\begin{center}
{\bf Appendix}
\end{center}

I. $\xi _0=\eta _0$ (circular orbit) 
\begin{equation}
\label{24} 
\begin{array}{c}
C_{mn_r}=\int \rho d\rho d\varphi \frac \alpha {\pi ^{1/2}}\exp [-\frac
12\xi _0^2-\frac 12(\xi ^2+\eta ^2)+\xi _0(\xi +i\eta )] \\ 
\times \left( \frac{n_r!\alpha ^2}{\pi (\left| m\right| +n_r)!}\right)
^{1/2}e^{-im\varphi }\tilde \rho ^{\left| m\right| }e^{-\frac 12\tilde \rho
^2}L_{n_r}^{\left| m\right| }(\tilde \rho ^2). 
\end{array}
\end{equation}
Using $\xi ^2+\eta ^2=\tilde \rho ^2,\xi +i\eta =\tilde \rho e^{i\varphi },$
and 
\begin{equation}
\label{25}\int_0^{2\pi }\exp [\xi _0\tilde \rho e^{i\varphi }]e^{im\varphi
}d\varphi =\left\{ 
\begin{array}{c}
2\pi 
\frac{(\xi _0\tilde \rho )^m}{m!},\ (m\geq 0)\ , \\ 0\qquad \quad ,\ (m<0)\
, 
\end{array}
\right. 
\end{equation}
we get 
\begin{equation}
\label{26}C_{mn_r}=\left[ \frac{n_r!}{(\left| m\right| +n_r)!}\right]
^{1/2}e^{-\xi _0^2/2}\frac{2\xi _0^m}{m!}\int_0^\infty d\tilde \rho \tilde
\rho ^{2\left| m\right| +1}e^{-\tilde \rho ^2}L_{n_r}^{\left| m\right|
}(\tilde \rho ^2)\ . 
\end{equation}
Using 
\begin{equation}
\label{27}2\int_0^\infty x^{2\lambda +1}e^{-x^2}L_n^\mu (x^2)dx=(-)^n\Gamma
(\lambda +1)\left( 
\begin{array}{c}
\lambda -\mu \\ 
n 
\end{array}
\right) , 
\end{equation}
it is seen that the integral in (26) does not vanish only for $n_r=0$, 
\begin{equation}
\label{28} 
\begin{array}{c}
C_{mn_r}=e^{-\xi _0^2/2}\left( \frac 1{m!}\right) ^{3/2}\cdot 2\xi _0^m
\frac 12m!\delta _{n_r0},(m\geq 0) \\ 
\quad =\left\{ 
\begin{array}{c}
\xi _0^me^{-\xi _0^2/2}\left( \frac 1{m!}\right) ^{1/2}\delta _{n_r0}\
,\quad (m\geq 0), \\ 
\qquad 0\qquad \ \quad \qquad \qquad ,\quad (m<0). 
\end{array}
\right. 
\end{array}
\end{equation}
II. $\xi _0\neq \eta _0$ (elliptic orbit) 
\begin{equation}
\label{29}C_{mn_r}=\left[ \frac{n_r!}{\pi ^2(\left| m\right| +n_r)!}\right]
^{1/2}e^{-\xi _0^2/2}\int \exp [\xi _0\xi +i\eta _0\eta ]e^{-im\varphi
}\tilde \rho ^{\left| m\right| +1}e^{-\tilde \rho ^2}L_{n_r}^{\left|
m\right| }(\tilde \rho ^2)d\tilde \rho d\varphi \ . 
\end{equation}
Using 
\begin{equation}
\label{30}\int_0^{2\pi }d\varphi \exp [\xi _0\xi +i\eta _0\eta
]e^{-im\varphi }=\left\{ 
\begin{array}{c}
2\pi \sum_{k=0}^\infty 
\frac{(A\tilde \rho )^k(B\tilde \rho )^{k+m}}{k!(k+m)!}\ ,\qquad (m\geq 0),
\\ 2\pi \sum_{k=0}^\infty \frac{(A\tilde \rho )^{k-m}(B\tilde \rho )^k}{%
k!(k-m)!}\ ,\qquad (m<0), 
\end{array}
\right. 
\end{equation}
(29) is reduced to (13).

\end{document}